\documentclass[11pt]{article}
\usepackage[totalwidth=460truept,totalheight=600truept]{geometry}
\usepackage{latexsym,amssymb,amsmath,graphicx}
\usepackage[T1]{fontenc}

\global\arraycolsep=1truept
\newcommand{\beq}{\begin{equation}}
\newcommand{\eeq}{\end{equation}}
\def\bea{\begin{eqnarray}}
\def\eea{\end{eqnarray}}
\begin{document}

\hfill UB-ECM-PF-09/24


\vskip 1.4truecm

\begin{center}
{\huge \textbf{Substituting fields within the action:}}

{\huge \textbf{\large \vskip .1truecm}}

{\huge \textbf{consistency issues and some applications}}

\vskip 1.5truecm

\textsl{Josep M. Pons}

\textit{Departament d'Estructura i Constituents de la Mat\`eria and Institut de
Ci\`encies del Cosmos, Facultat de F\'{\i}sica, Universitat de Barcelona, }

\textit{Diagonal 647, E-08028 Barcelona, Catalonia, Spain. }

pons@ecm.ub.es

\vskip 2truecm

\textbf{Abstract}
\end{center}

\bigskip

{\small In field theory, as well as in mechanics, the substitution of some fields in terms of other fields 
at the level of the action raises an issue of consistency with respect to the equations of motion. We 
discuss this issue and give an expression which neatly displays the difference between doing the 
substitution at the level of the Lagrangian or at the level of the equations of motion. Both operations do 
not commute in general. A very relevant exception is the case of auxiliary variables, which are discussed in 
detail together with some of their relevant applications. We discuss the conditions for the preservation of 
symmetries - Noether as well as non-Noether - under the reduction of degrees of freedom provided by the 
mechanism of substitution. We also examine how the gauge fixing procedures fit in our framework and give 
simple examples on the issue of consistency in this case.}

\vskip 1truecm

\vfill\eject

\section{Introduction}
In many instances of field theory, generally with the aim to make easier the treatment of the system, and in 
particular the obtention of solutions, some reduction procedures are sometimes introduced. A specific case 
is that of dimensional reduction, by which, as its name 
clearly indicates, the dimensions of spacetime are reduced and the system becomes simplified. Another type 
of reduction is based on the introduction of relations between the fields; this is the one we will be 
interested in here. At the level of the equations of motion (EOM), this reduction of degrees of freedom 
obviously amounts to the addition of new equations. Thus, the set of solutions of all the equations, old and 
new, if such a set exists, will be a subset of the solutions of the original theory. We say in such a case 
that the solutions of the new EOM are upliftable to solutions of the EOM of the original theory. 

One must be aware, though, that for theories whose dynamics is derived from a variational principle, 
problems may arise if one tries to implement these relations - which we will henceforth call constraints - 
at the level of the Lagrangian. In a nutshell: the processes of implementing the constraints at the level of 
the Lagrangian or at the level of the EOM do not commute in general. In this note we will discuss the 
consequences, at the level of the variational principle, of introducing constraints, and we will give a 
formula which explicitely shows the non-commutativity between both processes. Sometimes this commutativity 
is crucial for a reduction to make sense, because it will guarantee - at least in the classical setting here 
considered - the preservation of the physical content of the original theory. When commutatitvity holds, we 
say that the reduction is consistent. 

Once the conditions to guarantee the consistency of a reduction process - of the type described above - are 
properly understood, one can consider the inverse route in which, instead of a reduction, an enlargement of
the system is made, with the introduction of new variables, in such a way that the new system brings back 
the original one under a consistent reduction. It happens that sometimes the simplification for the 
treatment of the system is achieved not by reducing, but by enlarging it. A typical procedure to this effect 
is the introduction of auxiliary variables, to which we devote a section of this paper.

Aside from the aforementioned consistency issue, one can consider the different issue of preservation of 
symmetries under the substitution of some fields. These two different issues are often related because 
sometimes the reduction - or enlargement - of degrees of freedom is connected with symmetry considerations. 
We give a simple condition which ensures the preservation of continuous symmetries, with the geometric 
interpretation of being the requirement of tangency of the infinitesimal variations defining the symmetry to 
the constraint surface associated with the reduction. 

We study with some detail the reductions made by the elimination of auxiliary variables, and show that the 
preservation of continuous symmetries is guaranteed in this case, with the additional result that all 
Noether symmetries are preserved as such.

\vspace{4mm}

We will limit ourselves to constraints that can be expressed as the determination of some variables - that 
is, fields of field components - in terms of the rest, by way of a local functional - depending on the 
fields and their spacetime derivatives in a finite number. Our constraints are thus more general than the 
purely holonomic ones - which have no spacetime derivatives - but not the most general ones one can 
conceive. In Section 2 we prove our main result. In Section 3 the conditions for the preservation 
of symmetries under the reduction are discussed. Section 4 is devoted to auxiliary variables and some of 
their different physical impersonations. Section 5 deals with gauge fixing constraints and Section 6 is 
devoted to conclusions.

\section{Field substitutions}
\label{fieldsub}\setcounter{equation}{0}

Let us consider a field theory, governed by a variational principle with action 
\beq
{\cal S} = \int {\cal L}\,,
\label{act}
\eeq
where ${\cal L}$ is the Lagrangian density, with depends on the fields and their derivatives, in a finite 
number. Our aim is to study the consequences, as regards the dynamics and the action principle, of the 
substitution of some fields, let us call them $\psi$, by local functionals of the rest of the fields, which 
we denote $\phi$. Thus $\phi$ and $\psi$ represent a certain number of fields or field components, with 
their indices suppressed\footnote{An index free notation for the fields will be used throughout the paper.}. 

We set $\,\psi= F(\phi,\partial_\mu\phi, \partial_{\mu\nu}\phi,\ldots)\,$ to define, from the Lagrangian 
${\cal L}[\phi,\psi]$, the reduced Lagrangian ${\cal L}_{{}_r}:= {\cal L}_{|_{\psi\to F}}$, where 
$\psi\to F$ includes $\,\partial_{\mu}\psi\to \partial_{\mu}F$, etc. What is the effect of this substitution 
at the level of the variational principle? This is the issue which will be explored in the following.

Let us consider the variation of the action under arbitrary variations of the fields\footnote{Variations 
will always be ``active'', that is, they will never touch the spacetime coordinates.}. We have
$$
\delta\, {\cal S} = \int \delta{\cal L} = \int ([{\cal L}]_{{}_\phi}\delta \phi +[{\cal 
L}]_{{}_\psi}\delta \psi)+ {\rm b.t.}\,,
$$
where $[{\cal L}]_{{}_\phi}$ stands for the Euler-Lagrange functional
derivative of ${\cal L}$ with respect to $\phi$, etc., and ${\rm b.t.}$ 
represents generic boundary terms, that is, an integration on the boundary $\partial{\cal M}$ of the 
manifold ${\cal M}$ where the integration in (\ref{act}) takes place. Though important in other contexts, 
these boundary terms will play no role in our discussion.

One can define also the action for the reduced Lagrangian, ${\cal S}_{{}_r}= \int {\cal L}_{{}_r}$, and its 
variation 
\beq\delta\, {\cal S}_{{}_r} = \int \delta{\cal L}_{{}_r} = \int [{\cal L}_{{}_r}]_{{}_\phi}\delta \phi + 
{\rm b.t.}\,.
\label{delsr}
\eeq
It is easy to see that the following relation holds ($(\delta\, {\cal S})_{|_F}$ represents $(\delta\, {\cal 
S})_{|_{\psi\to F}}$ )
\beq(\delta\, {\cal S})_{|_{F}} = \delta\, {\cal S}_{{}_r}
\label{equal}
\eeq
because, associated with the process of reduction $\psi\to F$, the variations $\delta\psi$ must be 
understood as $(\delta\psi)_{|_{F}} = \delta F$ and thus (\ref{equal}) is nothing but the ordinary 
application of the chain rule. 

Let us now expand the first side of the equality (\ref{equal}) (in general the subscript $F$ represents the 
substitution $\psi\to F$ everywhere, including derivatives). 
\bea 
(\delta\, {\cal S})_{|_{F}} &=& \int \Big(([{\cal L}]_{{}_\phi})_{|_{F}}\delta \phi 
+([{\cal L}]_{{}_\psi})_{|_{F}}\delta F\Big)+ {\rm b.t.}\nonumber\\
&=& \int \Big(([{\cal L}]_{{}_\phi})_{|_{F}}\delta \phi +([{\cal L}]_{{}_\psi})_{|_{F}}(\frac{\partial 
F}{\partial \phi} \delta \phi+\frac{\partial F}{\partial\phi_{,\mu}}\delta 
\phi_{,\mu}+\frac{\partial F}{\partial\phi_{,\mu\nu}}\delta\phi_{,\mu\nu}+\ldots)\Big)+ {\rm b.t.}\nonumber\\
&=& \int \Big(
([{\cal L}]_{{}_\phi})_{|_{F}} + ([{\cal L}]_{{}_\psi})_{|_{F}}\frac{\partial F}{\partial \phi} - 
\partial_{\mu}(([{\cal L}]_{{}_\psi})_{|_{F}}\frac{\partial F}{\partial\phi_{,\mu}}) + 
\partial_{\mu\nu}(([{\cal L}]_{{}_\psi})_{|_{F}}\frac{\partial F}{\partial\phi_{,\mu\nu}})
+\ldots \Big)\,\delta \phi\nonumber\\
&+&{\rm b.t.}\,,
\label{secside}
\eea
where $\phi_{,\mu}:= \partial_{\mu}\phi$ are derivatives with respect to the coordinates of the manifold 
--in a given patch--, etc. Having no effect on the spacetime coordinates, the variations of the fields 
commute with the spacetime derivatives. Here and henceforth, dots as in the last equation represent obvious 
contributions from higher derivatives of the fields.

The second side of (\ref{equal}) has been expanded in (\ref{delsr}). Since (\ref{secside}) must be equal to 
(\ref{delsr}) and the variations $\delta \phi$ are arbitrary - they may even vanish outside a finite region 
of spacetime -, we obtain
\beq
\fbox{$ \displaystyle [{\cal L}_{{}_r}]_{{}_\phi} = 
([{\cal L}]_{{}_\phi})_{|_{F}} + ([{\cal L}]_{{}_\psi})_{|_{F}}\frac{\partial 
F}{\partial \phi} - \partial_{\mu}(([{\cal L}]_{{}_\psi})_{|_{F}}\frac{\partial F}{\partial\phi_{,\mu}}) + 
\partial_{\mu\nu}(([{\cal L}]_{{}_\psi})_{|_{F}}\frac{\partial F}{\partial\phi_{,\mu\nu}}) +\ldots
\label{theres}
$}
\eeq
Expression (\ref{theres}) displays in the right hand side, after the first term, what must be read
as the chain rule for the functional Euler-Lagrange derivative. It explicitely shows the difference, as 
regards the substitution $\psi\to F$, between 
doing it at the level of the Lagrangian, ${\cal L}\to{\cal L}_{{}_r}$, or at the level of the EOM. Clearly, 
both operations do not commute. If for $[{\cal L}]=0$ we symbolise all the EOM for ${\cal L}$, etc., we have 
indeed that $$[{\cal L}]_{|_{F}}=0\quad \Rightarrow\quad [{\cal L}_{{}_r}]=0\,,$$ 
but {\sl not} the other way around\footnote{Note that in the particular case in which $F$ does not depend on 
the fields $\phi$ (so $\psi\to F$ sets the fields $\psi$ to specific configurations), we obviously have 
$[{\cal L}_{{}_r}]_{{}_\phi} = 0 \Leftrightarrow ([{\cal L}]_{{}_\phi})_{|_{F}}=0$, but equations 
$[{\cal L}]_{{}_\psi})_{|_{F}}=0$ are missing.}. As a consequence, in general, a solution of the EOM for 
${\cal L}_{{}_r}$ is not upliftable (through the definition $\psi = F$) to a solution of the EOM for 
${\cal L}$.

In the case of mechanics a simplified version of (\ref{theres}), obtained by direct computation, appears in 
the appendix of \cite{Pons:1999az}. 

\vspace{4mm}

Some comments are now in order:

{\bf \underline{Comment 1}}

The dynamics generated by ${\cal L}_{{}_r}$ is equivalent to the dynamics obtained by applying to the 
original Lagrangian the Lagrangian multipliers method. This is a field theory generalization of the results, 
see for instance  \cite{Lanczos}, in mechanical systems. Indeed, defining the enlarged Lagrangian with the 
Lagrange multipliers\footnote{In our index free notation, there are as many Lagrange multipliers as substitutions $\psi \to F$.} $\lambda$, ${\cal L}_{{}_e} :=
 {\cal L} + \lambda(\psi - F) $, its EOM are
\beq
[{\cal L}_{{}_e}]_{{}_\phi} = [{\cal L}]_{{}_\phi} - \lambda\frac{\partial F}{\partial \phi} + 
\partial_\mu(\lambda\frac{\partial F}{\partial\phi_{,\mu}}) + \ldots =0
\eeq
\beq
[{\cal L}_{{}_e}]_{{}_\psi} =
[{\cal L}]_{{}_\psi} + \lambda = 0 
\eeq
\beq
[{\cal L}_{{}_e}]_{{}_\lambda} = \psi - F=0\,,
\eeq
and plugging the second and third EOM into the first we arrive, in view of (\ref{theres}), at the EOM for 
the reduced Lagrangian, $[{\cal L}_{{}_r}]_{{}_\phi}=0$.

We may notice that the variables $\psi,\lambda$ play the role of auxiliary variables in ${\cal L}_{{}_e}$, 
which is the subject of section \ref{auxdef}.

\vspace{4mm}

\underline{\bf Comment 2}

Expression (\ref{theres}) bears a strong resemblance in structure with a formula obtained in 
\cite{Pons:2009nb}, which will be used later on,
\beq 
\delta [{\cal L}]_{{}_A}=[\delta{\cal L}]_{{}_A} -[{\cal L}]_{{}_B}\frac{\partial\,\delta 
\varphi^{{}_B}}{\partial \varphi^{{}_A}}
 +\partial_{\mu}([{\cal L}]_{{}_B}\frac{\partial\,\delta 
\varphi^{{}_B}}{\partial\varphi^{{}_A}_{{}_{,\mu}}})-\partial_{\mu\nu}([{\cal
L}]_{{}_B}\frac{\partial\,\delta \varphi^{{}_B}}{\partial\varphi^{{}_A}_{{}_{,\mu\nu}}})+\ldots\,,
\label{kiyoshi}
\eeq
where $\varphi^{{}_A},\,\varphi^{{}_B}$ represent any field of field component. This equation was obtained 
from considerations concerning the variation of the Euler-Lagrange derivatives of a given Lagrangian versus 
the Euler-Lagrange derivatives of the variation of this Lagrangian. Formula (\ref{kiyoshi}) is valid for 
variations which are local functionals of the fields. Again, it all boils down to the chain rule for the 
functional derivative.

\vspace{4mm}

\underline{\bf Comment 3}

The Noether identities for gauge theories can be quickly derived from this 
chain rule: consider an action ${\cal S} = \int {\cal L}$, functional of the fields $\varphi$, and suppose 
that $\delta \varphi$ is an infinitesimal Noether gauge symmetry, depending on some arbitrary infinitesimal 
function $\epsilon(x)$ and its spacetime derivatives up to a certain order. An infinitesimal parameter 
$\delta \lambda$ is present in the function $\epsilon$; since $\delta \varphi$ is first order in this 
infinitesimal parameter, the dependence of $\delta \varphi$ on $\epsilon$ must be linear,
\beq
\delta \varphi = R_{{}_\varphi} \epsilon + R_{{}_\varphi}^\mu \epsilon_{,\mu} + R_{{}_\varphi}^{\mu\nu} 
\epsilon_{,\mu\nu} + \ldots,
\label{deltlin}\eeq 
for some functions $R_{{}_\varphi},\, R_{{}_\varphi}^\mu,\,R_{{}_\varphi}^{\mu\nu},\cdots$ of the fields and 
their spacetime derivatives. We have
$$\delta\, {\cal S} = \int \delta{\cal L} = \int [{\cal L}]_{{}_\varphi}\delta \varphi 
+ {\rm b.t.} = {\rm b.t.}\,,
$$
(where the last equality comes from the Noether condition of symmetry) and the computation of the 
functional derivative of $\delta\, {\cal S}$ with respect to $\epsilon(x)$ gives 
$$\frac{\delta}{\delta\epsilon}(\delta\, {\cal S})  = [{\cal L}]_{{}_\varphi}\frac{\partial\,\delta 
\varphi}{\partial \epsilon} 
- \partial_{\mu}([{\cal L}]_{{}_\varphi}\frac{\partial\,\delta 
\varphi}{\partial\, \epsilon_{{}_{,\mu}}}) + 
\partial_{\mu\nu}([{\cal L}]_{{}_\varphi}\frac{\partial\,\delta 
\varphi}{\partial\, \epsilon_{{}_{,\mu\nu}}}) +\ldots\,,
$$
(in fact we could cancel out the infinitesimal parameter $\delta \lambda$ form both sides) but since 
$\delta\, {\cal S}$ is a boundary term, $\frac{\delta}{\delta\epsilon}(\delta\, {\cal S})$ must 
vanish for values of $x$ in $\epsilon(x)$ denoting points in the bulk of the manifold, that is, not in the 
boundary. We infer, using (\ref{deltlin}), that
\beq
 [{\cal L}]_{{}_\varphi} R_{{}_\varphi} - \partial_\mu([{\cal L}]_{{}_\varphi}R_{{}_\varphi}^\mu)  +  
\partial_{\mu\nu}( [{\cal L}]_{{}_\varphi} R_{{}_\varphi}^{\mu\nu} )   +\ldots   = 0\,,
\label{noet-id}\eeq
which is the Noether identity for the gauge symmetry $\delta \varphi$. Thus we realize that the Noether 
identity is just the expression of the independence of $\delta\, {\cal S}$ with respect to the arbitrary 
function $\epsilon$ present in the variation $\delta$.

\section{Preservation of symmetries}
\label{preserv}\setcounter{equation}{0}
In the computations leading to (\ref{theres}), the variations $\delta \phi, 
\,\delta \psi$ were arbitrary and the relation $(\delta\psi)_{|_{F}} = \delta F$ was just the expression of 
the substitution $\psi\to F$ at the level of the variations themselves. A completely different matter occurs 
when the variations represent infinitesimal symmetries and are given by specific functionals of the fields 
(as is the case in (\ref{kiyoshi})). 
Without changing the notation, in this section $\delta \phi, \,\delta \psi$ stand for specific infinitesimal 
local functionals of the fields\footnote{The infinitesimality can always be associated, for each independent 
symmetry, with the presence of an infinitesimal parameter as a global factor in all 
$\delta \phi, \,\delta \psi$.}. In this case it is clear that $(\delta\psi)_{|_{F}}$ is not necessarily 
equal to $(\delta F)_{|_{F}}$\footnote{Note that we write now $(\delta F)_{|_{F}}$ because $\delta\phi$ may 
depend on $\psi$ or on its derivatives.}, and that only under very restricted circumstances the equality 
will hold. Indeed we may define the symmetry to be consistent with the substitution $\psi\to F$ if and 
only if $(\delta\psi - \delta F)_{|_{F}} =0$. In such case, with the general definition 
\beq
\delta_{{}_{r}}\phi:=(\delta\phi)_{|_{F}}\,,
\label{defdeltared}
\eeq 
we still have then the crucial result that 
$(\delta\, {\cal S})_{|_{F}} = \delta_{{}_{r}}\, {\cal S}_{{}_r}$ as an application of the ordinary chain 
rule. In particular this means that if  $\delta\, {\cal S}$ is a boundary term, so it is $\delta_{{}_r}\, 
{\cal S}_{{}_r}$. This allows us to formulate the following:

\underline{\bf Proposition 1.-}

{\sl If $\delta \phi, \,\delta \psi$ is an infinitesimal Noether symmetry\footnote{Noether symmetries are 
characterized by the fact that $\delta{\cal L}$ is a divergence, or, what is the same, $\delta{\cal S}$ is a 
boundary term.} for ${\cal L}$, consistent with the substitution $\psi\to F$, that is, if the 
\beq
{Tangency\ condition:}\qquad \Big(\delta(\psi - F)\Big)_{|_{F}} =0\,,\qquad\qquad\qquad\qquad\qquad
\label{conssymm}
\eeq
holds, then $\delta_{{}_r}\phi$ is a Noether symmetry for ${\cal L}_{{}_r}$.}

\vspace{4mm}

This result is general for Noether symmetries, either rigid or gauge. The obvious geometric interpretation 
of (\ref{conssymm}) is that of a tangency condition of the infinitesimal symmetry transformation with 
respect to the constraint surface defined by $\psi - F=0$. Note that this condition 
(\ref{conssymm}) for the preservation of the Noether symmetries is independent of the analysis made in the 
previous section on consistent reductions. These are two independent issues. A case where a substitution - 
be it consistent or not - of this type, $\psi\to F$, is performed, with the aim of getting a reduced theory 
ensuring the preservation of certain symmetries is known in the literature as the inverse Higgs mechanism 
\cite{Ogi}, see \cite{Gomis:2008jc} for a recent application.

\vspace{4mm}

For purposes that will become clear in the development of the next section, let us elaborate with more 
detail on our findings concerning the preservation of the Noether symmetries. With $\delta \phi, \,\delta 
\psi$ being again specific variations, that is, functionals of the fields, we have
$$
\delta{\cal L} = [{\cal L}]_{{}_\psi}\delta\psi + [{\cal L}]_{{}_\phi}\delta\phi + {\rm div.}\,,
$$
where by ``div.'' we mean generic divergences. Then, using equation (\ref{theres}) and the definition 
(\ref{defdeltared}) in the second equality, 
\bea
\hspace{-1mm}
&&(\delta{\cal L})_{|_{F}} = ([{\cal L}]_{{}_\psi})_{|_{F}}(\delta\psi)_{|_{F}} 
+ ([{\cal L}]_{{}_\phi})_{|_{F}}(\delta\phi)_{|_{F}} + {\rm div.}\nonumber\\ 
&&=([{\cal L}]_{{}_\psi})_{|_{F}}(\delta\psi)_{|_{F}} +\Big([{\cal L}_{{}_r}]_{{}_\phi} 
-([{\cal L}]_{{}_\psi})_{|_{F}}\frac{\partial F}{\partial \phi}
+\partial_{\mu}(([{\cal L}]_{{}_\psi})_{|_{F}}\frac{\partial F}{\partial\phi_{,\mu}})
-\partial_{\mu\nu}(([{\cal L}]_{{}_\psi})_{|_{F}}\frac{\partial F}{\partial\phi_{,\mu\nu}}) 
+\ldots\Big)\delta_{{}_r}\phi+ {\rm div.}\nonumber\\ 
&&=([{\cal L}]_{{}_\psi})_{|_{F}}(\delta\psi)_{|_{F}} +[{\cal L}_{{}_r}]_{{}_\phi}\delta_{{}_r}\phi
- ([{\cal L}]_{{}_\psi})_{|_{F}}\frac{\partial F}{\partial \phi}\delta_{{}_r}\phi
-([{\cal L}]_{{}_\psi})_{|_{F}}\frac{\partial F}{\partial\phi_{,\mu}}\delta_{{}_r}\phi_{,\mu} 
-([{\cal L}]_{{}_\psi})_{|_{F}}\frac{\partial F}{\partial\phi_{,\mu\nu}}\delta_{{}_r}\phi_{,\mu\nu} + {\rm 
div.} \nonumber\\ &&= ([{\cal L}]_{{}_\psi})_{|_{F}}(\delta\psi)_{|_{F}} 
+[{\cal L}_{{}_r}]_{{}_\phi}\delta_{{}_r}\phi
-([{\cal L}]_{{}_\psi})_{|_{F}}\delta_{{}_r} F = [{\cal L}_{{}_r}]_{{}_\phi}\delta_{{}_r}\phi 
+ ([{\cal L}]_{{}_\psi})_{|_{F}}(\delta(\psi - F))_{|_{F}}+ {\rm div.}\nonumber\\ 
&&= \delta_{{}_r}{\cal L}_{{}_r} + ([{\cal L}]_{{}_\psi})_{|_{F}}(\delta(\psi - F))_{|_{F}}+ {\rm div.}\,.
\eea
We have obtained an interesting equation
\beq\fbox{$ \displaystyle (\delta{\cal L})_{|_{F}}= \delta_{{}_r}{\cal L}_{{}_r} 
+ ([{\cal L}]_{{}_\psi})_{|_{F}}(\delta(\psi - F))_{|_{F}}+ {\rm div.}
\label{noet-cond}
$}\eeq
which informs us of some sufficient conditions for the preservation of Noether symmetries under the 
substitution $\psi \to F$. A first application of (\ref{noet-cond}) is immediate: noticing that if 
$\delta{\cal L}$ is a divergence so it is $(\delta{\cal L})_{|_{F}}$, we obtain from 
(\ref{noet-cond})
\beq
(\delta(\psi - F))_{|_{F}}=0\quad\Rightarrow\quad \Big(\delta{\cal L}= {\rm div.}\ \Rightarrow\ 
\delta_{{}_r}{\cal L}_{{}_r}= {\rm div.}\Big)
\label{noet-to-noet}
\eeq
which is the result already stated in {\bf Proposition 1}. But now (\ref{noet-cond}) is ready for yet 
another, second application, to be made in the next section.

\section{Auxiliary variables}
\setcounter{equation}{0}
\label{auxdef}

Suppose that the variables $\psi$ are auxiliary variables. By this we mean that they can be isolated in 
terms of the other variables by using {\sl their own} EOM. In this case we can take $F$ such that
\beq
[{\cal L}]_{{}_\psi}=0\quad \Longleftrightarrow\quad \psi= F(\phi,\partial_\mu\phi, 
\partial_{\mu\nu}\phi,\ldots)\,. 
\label{aux}\eeq
In such a particular case, $([{\cal L}]_{{}_\psi})_{|_{F}}$ vanishes indentically, and (\ref{theres}) 
becomes $[{\cal L}_{{}_r}]_{{}_\phi} = ([{\cal L}]_{{}_\phi})_{|_{F}}$. This proves that when auxiliary 
variables are substituted back into the Lagrangian by using their own EOM, the dynamics for 
the rest of the variables remains unaltered, that is
\beq  
{\rm Elimination\ of\ auxiliary\ variables:} \qquad [{\cal L}]_{|_{F}}= 0 \quad 
\Longleftrightarrow\quad [{\cal L}_{{}_r}] = 0\,.\qquad\qquad\qquad\qquad
\label{auxequiv}
\eeq
This is a very special case in which both procedures, i.e., substitution into the Lagrangian or substitution 
into the EOM, commute. Whether this result may hold in particular cases for substitutions not coming from 
auxiliary variables depends on the specifics of each situation (for instance, in the case that 
$[{\cal L}]_{{}_\phi}=0\ \Rightarrow\ [{\cal L}]_{{}_\psi}=0$ it seems likely that equivalence may hold), 
but it must be checked on a case by case basis.
\subsection{Auxiliary variables and preservation of symmetries}
\label{auxpres}

Expression (\ref{noet-cond}) already contains the proof that the reductions made by the elimination of 
auxiliary variables preserve the Noether symmetries. In fact one derives from (\ref{noet-cond})
\beq
([{\cal L}]_{{}_\psi})_{|_{F}}=0\quad\Rightarrow\quad \Big(\delta{\cal L}= {\rm div.}\ \Rightarrow\ 
\delta_{{}_r}{\cal L}_{{}_r}= {\rm div.}\Big)\,.
\label{noet-to-noet-aux}
\eeq
Thus the preservation of Noether symmetries is always guaranteed in this case.

\vspace{4mm}
 
One may ask nonetheless whether the tangency condition holds for Noether symmetries in reductions 
driven by auxiliary variables. In a strict sense, the answer in general is in the negative, even though it 
still holds on shell. Let us prove it. Since (\ref{aux}) is satisfied, we see that the condition 
(\ref{conssymm}) can be equivalently written as:
\beq 
{\rm Tangency\ condition\ with\ auxiliary\ variables:}\qquad 
(\delta [{\cal L}]_{{}_\psi})_{|_F}=0\,,\qquad\qquad \qquad \qquad  
\label{aux2}
\eeq
where now the subscript $F$ has the equivalent meaning of requiring $[{\cal L}]_{{}_\psi}=0$ with $\psi$
being the auxiliary fields. Our next task is to check whether (\ref{aux2}) is satisfied. According to 
(\ref{kiyoshi}), we know that
\beq 
\delta [{\cal L}]_{{}_\psi}=[\delta{\cal L}]_{{}_\psi} -[{\cal L}]_{{}_\varphi}\frac{\partial\,\delta 
\varphi}{\partial \psi}
+
\partial_{\mu}([{\cal L}]_{{}_\varphi}\frac{\partial\,\delta 
\varphi}{\partial\psi_{,\mu}})-\partial_{\mu\nu}([{\cal
L}]_{{}_\varphi}\frac{\partial\,\delta \varphi}{\partial\psi_{,\mu\nu}})+\ldots\,,
\label{kiyoshi2}
\eeq
where $\varphi$ represents any field, $\varphi= \phi,\,\psi$. Note that (\ref{kiyoshi2}) gets simplified 
when we take into account the assumption that the variations are a Noether symmetry for the original 
Lagrangian ${\cal L}$. This implies in particular that $[\delta{\cal L}]_{{}_\psi}=0$. Since $\psi$ are 
auxiliary variables, the satisfaction of (\ref{aux}) guarantees that (\ref{theres}) is just 
$[{\cal L}_{{}_r}]_{{}_\phi} = ([{\cal L}]_{{}_\phi})_{|_{F}}$. 
Then equation (\ref{kiyoshi2}) becomes, under $\psi\to F$, 
\beq 
(\delta [{\cal L}]_{{}_\psi})_{|_{F}}=-[{\cal L}_{{}_r}]_{{}_\phi}\frac{\partial\,\delta \phi}{\partial 
\psi}|_{{}_F}  +
\partial_{\mu}\Big([{\cal L}_{{}_r}]_{{}_\phi}\frac{\partial\,\delta 
\phi}{\partial\psi_{,\mu}}|_{{}_F}\Big)-\partial_{\mu\nu}\Big([{\cal 
L}_{{}_r}]_{{}_\phi}\frac{\partial\,\delta \phi}{\partial\psi_{,\mu\nu}}|_{{}_F}\Big)+\ldots\,,
\label{kiyoshi3}
\eeq
and the obstruction for the satisfaction of (\ref{aux2}) is identified: in general (\ref{aux2}) 
will not be true as long as the variations $\delta \phi$ functionally depend on the auxiliary variables.

Note however that the eventual violation of (\ref{aux2}) is mild, for it is still satisfied on 
shell, that is, for $[{\cal L}_{{}_r}]_{{}_\phi}=0$. Thus in the case of auxiliary variables a weak form of 
the tangency condition still holds for Noether symmetries. 

\vspace{4mm}

Finally, let us explore the fate of the continuous non-Noether - or on shell - symmetries. They only need to 
satisfy the requirement of mapping solutions into solutions and are characterized by the property 
(see \cite{Pons:2009nb})
\beq
{\rm On\ shell\ continuous\ symmetry:}\qquad (\delta[{\cal L}])_{{}_{[{\cal L}] = 0}} = 0\,. 
\label{on-shell-symm}
\eeq 
Here we prove that an on shell symmetry is preserved under the elimination of auxiliary variables. First 
note that, just by definition of auxiliary variables,
$$  
[{\cal L}] = 0\quad  \Longleftrightarrow\quad  \psi - F=0,\quad [{\cal L}_{{}_r}] = 0\,,
$$
which implies
$$ 
(\delta[{\cal L}])_{{}_{[{\cal L}] = 0}} = 0 \quad  \Longleftrightarrow\quad
\Big(\delta(\psi - F)\Big)_{{}_{\psi - F=0,\ [{\cal L}_{{}_r}] = 0}} =0,\quad (\delta[{\cal 
L}_{{}_r}])_{{}_{\psi - F=0,\ [{\cal L}_{{}_r}] = 0}}=0\,.
$$ 
The first term in the right hand side is the on shell - or weak - tangency condition, which is a result we 
already knew for Noether symmetries. As for the second term, using the definition (\ref{defdeltared}), 
it becomes
$$ 
(\delta[{\cal L}_{{}_r}])_{{}_{\psi - F=0,\ [{\cal 
L}_{{}_r}] = 0}}\quad = \quad (\delta_{{}_r}[{\cal L}_{{}_r}])_{{}_{[{\cal L}_{{}_r}] = 0}}\,,
$$ 
and therefore
\beq 
(\delta[{\cal L}])_{{}_{[{\cal L}] = 0}} = 0 \quad  \Longrightarrow\quad (\delta_{{}_r}[{\cal 
L}_{{}_r}])_{{}_{[{\cal L}_{{}_r}] = 0}}= 0\,.
\label{onshell}
\eeq
In view of the characterization (\ref{on-shell-symm}) of on shell continuous symmetries, this result 
(\ref{onshell}) allows us to conclude that $\delta_{{}_r}$ defines an on shell symmetry of 
${\cal L}_{{}_r}$ if $\delta$ is an on shell symmetry of ${\cal L}$.

\vspace{4mm}

In conclusion, we have proved the following:

\underline{\bf Proposition 2.-}

{\sl The original continuous symmetries are preserved through reductions made by the 
elimination of auxiliary variables. If the symmetries are of the Noether type, they will also be for the 
reduced theory.}
\subsection{An example}

Consider the Lagrangian in mechanics, extracted form \cite{Gomis:2008jc},
$$L = m({\vec v}\dot{\vec x} - \frac {1}{2}\vec v^2) + \kappa \frac {1}{2} \epsilon_{ij}v^i \dot v^j,
$$
where $\vec x=(x^1,\,x^2),\ \vec v=(v^1,\,v^2) $ are independent configuration variables. 
It is clear that the variables $v^i$ are auxiliary only for $ \kappa=0$, because the EOM then dictate 
$\vec v= \dot{\vec x}$. For $ \kappa\neq 0$ they are not auxilary anymore, but we can just keep the 
substitution $\vec v\to \dot{\vec x}$ in order to check the formula (\ref{theres}). $L_{{}_r}$ becomes
$$L_{{}_r} = \frac {1}{2}m (\dot{\vec x})^2  + \kappa \frac {1}{2} \epsilon_{ij}\dot x^i \ddot x^j\,.
$$
Now, $\ ([L]_{{}_{x^i}})_{|_{F}} = - m \ddot x^i$, $\ ([L]_{{}_{v^i}})_{|_{F}} = \kappa 
\epsilon_{ij}\ddot x^j\ $,  and 
$\ [L_{{}_r}]_{{}_{x^i}} = -m \ddot x^i - \kappa \epsilon_{ij} \dddot x^j$. One can immediately verify 
(\ref{theres}).

Note that the equivalence (\ref{auxequiv}) is only achieved for $\kappa=0$. Indeed, keeping always 
$m\neq 0$, the right hand side of (\ref{auxequiv}) is just $ \ddot x^i =0$ whereas the left hand side is 
$m \ddot x^i + \kappa \epsilon_{ij} \dddot x^j=0$. 

Note also that the substitution $\vec v\to \dot{\vec x}$ is consistent with the $SO(2)$ invariance present 
in $L$. In consequence, as discussed in section \ref{preserv}, $L_{{}_r}$ inherits this invariance, in 
this case, as a Noether symmetry.
\subsection{Making the Lagrangian polynomial: the string}
\label{first}
An obvious and well known example of the relevance of the auxiliary variables is that of the reformulation 
by \cite{Brink:1976sc,Deser:1976rb} of the Nambu-Goto action 
\cite{Nambu,Goto:1971ce} for the string, by intoducing the metric on the worldsheet as an independent 
field, whose components are auxiliary variables. It is straighforwardly extended to $p$-branes with the 
Lagrangian 
$$ {\cal L} = \sqrt{-g}\Big(g^{\mu\nu}\partial_\mu X^{\!A} \partial_\nu X^{\!B} G_{\!A B}(X) -p +1\Big)\,,
$$
where $G_{\!A B}$ is the target space metric, $g_{\mu\nu}$ is the worldsheet metric, $g$ its determinant 
and $g^{\mu\nu}$ its inverse. The EOM for $g^{\mu\nu}$ are, for $p\neq 1$, 
$ g_{\mu\nu} - \partial_\mu X^{\!A} \partial_\nu X^{\!B} G_{\!A B} =0$, which makes $g_{\mu\nu}$ auxiliary 
variables. The string case, $p=1$, requires more care, because the EOM for the worldsheet metric, 
$$
g_{\mu\nu} - \frac{2}{g^{\rho\sigma}\partial_\rho X^{\!C} \partial_\sigma X^{\!D} G_{\!C D}}\partial_\mu 
X^{\!A}\partial_\nu X^{\!B} G_{\!A B} =0\,,
$$
does not really allow for the determination of $g_{\mu\nu}$. This is actually a consequence of the Weyl 
invariance for the two dimensional worldsheet. Note however that the substitution $ g_{\mu\nu} \to 
F_{\!\mu\nu}$ with $F_{\!\mu\nu} = f(x) \partial_\mu X^{\!A} \partial_\nu X^{\!B} G_{A B}$, for any 
arbitrary nonvanishing function $f$ on the worldsheet ($x$ represents the worldsheet coordinates), already 
implies $([{\cal L} ]_g)_{|_{F}}=0$, which is all that matters in order to guarantee the consistency of the 
reduction in this case. Note in addition that this substitution includes, in the selection of a specific 
function $f$, a gauge fixing for the Weyl invariance; this is in agreement with the fact that this 
invariance has no room for it to be realized in the reduced theory. There is no contradiction with the 
results obtained in section \ref{auxpres} because, striclty speaking, in the case $p=1$, according to our 
definition in section \ref{auxdef}, the fields $g_{\mu\nu}$ are not auxiliary, for they can not be isolated 
by the use of their own EOM. But they come close (they are auxiliary fields after the gauge fixing), and in 
a more loose sense we can still call them auxiliary variables.

The use of these auxiliary variables for the string and $p$-branes allows to circumvent the problems 
associated with the quantization of non-polynomial Lagrangians.  
\subsection{Reducing the order of the EOM: the example of $f(\mathcal R)$ gravity}
\label{second}
Another possible advantage of the mechanism of enlarging the field content of a theory through the 
addition of some auxiliary variables is the reduction of the order in derivatives 
of the EOM. An interesting example in this respect is provided by some modified theories of gravity. 
Consider for instance the Lagrangian for $f(\mathcal R)$ gravity 
\beq
{\cal L}
 = \sqrt{-g}\,f(\mathcal R)
   + {\cal L}_\mathrm m[\Psi,g_{\mu\nu}]\,,
\label{actionfr}
\eeq
where we have set $16\pi G=1$\,. In (\ref{actionfr}) $g$ is the determinant of the metric, $\mathcal R$ is 
the scalar curvature, and $\Psi$ denotes some matter fields minimally coupled to the metric.
Since (\ref{actionfr}) contains second derivatives of $g_{\mu\nu}$ the EOM will in general be fourth-order 
differential field equations. One can avoid this complication by following the ideas of  
\cite{Teyssandier:1983zz}. Let us introduce a couple of scalar fields as auxiliary variables, see for 
instance \cite{Nojiri:2006ri,Deruelle:2009pu}, $\lambda,\rho$, as follows. 
\beq
{\cal L}_{{}_{enl}} = \sqrt{-g}\,
   \Big(f(\rho) - \lambda\,(\rho - \mathcal R)\Big)
   + {\cal L}_\mathrm m[\Psi,g_{\mu\nu}]\,.
\eeq
Notice in fact that $\lambda$ is a Lagrange multiplier set to enforce the constraint $\rho - \mathcal R=0$, 
a procedure that should be familar to us after Comment 1 in section \ref{fieldsub}.
Actually, if $f''(\rho)\neq 0$, the variable $\rho$, taken alone, is auxiliary - instead, 
$\lambda$ alone is not. It can be isolated from its own EOM as $\rho=h(\lambda)$, with $h=(f')^{-1}$, and 
can be plugged into $ {\cal L}_{{}_{enl}}$, leaving $\lambda$ as a variable in the Lagrangian. We obtain 
$$
{\tilde {\cal L}}= \sqrt{-g}\,
   \Big(f(h(\lambda)) - \lambda\,(h(\lambda) - \mathcal R)\Big)
   + {\cal L}_\mathrm m[\Psi,g_{\mu\nu}]\,,
$$
which is Brans-Dicke theory \cite{Brans:1961sx} in the Jordan frame - with Brans-Dicke parameter
$\omega_0 = 0$ and some potential for the scalar field. 
Under the condition $h'(\lambda)\neq 0$, which is nothing but $f''(\rho)\neq 0$, $\lambda$ is now an 
auxiliary variable. Let us  
redefine $\lambda = e^{\chi}$. The conformal - Weyl - transformation $g_{\mu\nu}\to e^{-\chi}g_{\mu\nu}$ 
produces the Lagrangian in the Einstein frame,
$${\cal L}_{{}_E} = \sqrt{-g}\,\Big( {\mathcal R} - \frac{3}{2} g^{\mu\nu} 
\partial_\mu\chi\partial_\nu\chi - e^{-\chi}h(e^{\chi}) + e^{-2\,\chi}f(h(e^{\chi}))\Big)+ 
{\cal L}_\mathrm m[\Psi,e^{-\chi}g_{\mu\nu}]\,.
$$
At this point, the status of $\chi$ as an auxiliary variable is lost, and it becomes a 
dynamical variable. It might seem as if a new dynamical field has appeared out of the blue, but 
one should notice that it accounts for the reduction of the order of the EOM. 
Note that the coupling of the matter fields with the new metric is no longer the minimal one. See 
\cite{Nojiri:2006ri} for further details, references and discussion on the physical interpretation. 
\subsection{Closing the algebra of generators: supersymmetry and BRST symmetry} 
\label{third}
Auxiliary variables are used in supersymmetry as a means to obtain a closed algebra of the supersymmetry 
generators. Indeed, in a general theory, the algebra of generators of Noether symmetries may exhibit, in its 
right hand side, trivial Noether generators, made up with antisymmetric combinations of the EOM (see for 
instance \cite{Henneaux:1992ig}, chapter three). This defines the case of open algebras, which abound in 
supersymmetric theories unless auxiliary variables are introduced. In fact, such type of variables appears 
in a natural way in the superspace formulation, see for instance \cite{Freund:1986ws}. Thanks to them, the 
on shell matching of Bose and Fermi degrees of freedom can be extended off shell. 

\vspace{4mm}

Auxiliary variables play a similar role in the formulation of BRST symmetry \cite{Becchi:1974md} , which is 
the offspring of a former gauge symmetry after it has been gauge 
fixed at the level of the action. A quadratic term of the type
$\sum_{a=1}^{n} (f^a)^2$ in the Lagrangian may be replaced as 
$$
\sum_{a=1}^{n} (f^a)^2\quad \to\quad  2\,\sum_{a=1}^{n}(B^a f^a) - \sum_{a=1}^{n}(B^a)^2\,,
$$ 
where $ B^a$ are the Nakanishi-Lautrup auxiliary fields \cite{nak,laut}. By integrating them out in a path 
integral formulation - which in this quadratic case is equivalent to the substitution of their own classical 
EOM -, the original quadratic term is recovered. The off shell nilpotent BRST Noether charge is constructed 
with the aid of these auxiliary fields. Nilpotency only holds on shell if the auxiliary fields are 
eliminated. 

In a more general case, in the context of the field-antifield formalism for gauge theories (see 
\cite{Gomis:1994he} for a review and references), one can prove \cite{Henneaux:1990ua} the equivalence of 
the path integrals before and after the elimination of the auxiliary variables.

\vspace{4mm}

The auxiliary variables count as off shell degrees of freedom, but do not count as on shell, that is, as 
physical degrees of freedom. In fact, using the techniques of the theory of constrained systems - also 
considered in the next section - as developed by Rosenfeld, Dirac and Bergmann (RDB), 
\cite{rosenfeld30,bergmann49a,bergbrun49,bergm3,dirac50,dirac4}\footnote{See \cite{Pons:2004pp} for a 
brief introduction to the RDB theory. References of books include
\cite{Sundermeyer:1982gv,Henneaux:1992ig,Gitman:1990qh}.}, a simple analysis in phase space of the BRST 
example just mentioned shows that the variables $\pi_a$, canonically conjugate to $B^a$, are primary 
constraints, and their stabilization yields $B^a - f^a$ as secondary constraints, which are the Lagrangian 
EOM for $B^a$. All together form a set of second class constraints which can be eliminated. In this case the 
Dirac bracket is nothing but the Poisson bracket for the original variables.
\section{Gauge fixing constraints}
\setcounter{equation}{0}
Another interesting case to consider is the implementation of a gauge fixing constraint at the level of the 
Lagrangian in a gauge theory. In this case one can use other tools, complementing formula (\ref{theres}), to 
analyze the issue. Using the RDB formalism, it was shown in \cite{Pons:1995ss}, appendix C of 
\cite{Pons:1998tt}, and \cite{Pons:2004ky}, that the effect of plugging the gauge fixing constraint into the 
Lagrangian can be compensated by adding to the equations of motion for the reduced theory some constraints 
that have disappeared as such along the process. Consider, as an example, pure electromagnetism in the 
temporal gauge $A_0 =0$ ($A_0$ is the time component of the gauge field). The reduced theory will miss the 
Gauss constraint, but once this constraint is imposed on the equations of motion for the reduced theory, 
equivalence is reached with the EOM of the original theory supplemented with the gauge fixing 
constraint\footnote{See \cite{Morchio:2009sy} for a recent examination of this issue in quantum field 
theory.}.

\vspace{4mm}

A formal example from mechanics may help to clarify why it is so. We can still use formula (\ref{theres}) 
to see what happens in this simple case. Consider a Lagrangian such that it does not
depend on the velocity of some variable $q$. The momenta associated with such variable
is, in the language of RDB, a primary constraint in phase space, because $ p=\frac{\partial L}{\partial \dot 
q} =0$ for this specific variable. Now apply the RDB techniques. If as a result of the phase space 
constraint analysis it turns out that $p$ 
is eventually first class, this means that the theory exhibits gauge freedom. A good gauge fixing constraint 
is then $q= c$, where $c$ is a constant, which converts $p=0$ into a second class constraint. Note that the 
EOM for $q$, $[L]_{{}_q}   =   \frac{\partial L}{\partial  q} - \frac{d}{d\,t}\frac{\partial L}{\partial 
\dot q}=0$, implies that in phase space we have the equation $\frac{d\,p}{d\,t}  = \frac{\partial 
L}{\partial  q}$, but since $\frac{d\,p}{d\,t}  =0$ as a result of the stabilization of the primary 
constraint, we end up with $\chi:=\frac{\partial L}{\partial  q} =0$ as a new constraint. In fact,  $\chi$ 
is an obvious constraint in the Lagrangian formalism, regardless of any phase 
space analysis, but we wanted to make the point that $\chi$ is essentially a secondary 
constraint in phase space, although it is written here with configuration-velocity variables. In any case, 
this constraint is all we need. Now we can apply (\ref{theres}) to the present situation. 
$([{\cal L}]_{{}_\psi})_{|_{F}}$ is $([L]_{{}_q})_{|_{q=c}} = \frac{\partial L}{\partial  
q}_{|_{q=c}} = \chi_{{}_c}$, where $\chi_{{}_c} := \chi_{|_{q=c}}$. Thus in this example we obtain the 
equivalence
$$
[L]_{|_{q= c}}=0\quad  \Leftrightarrow\quad  [L_{{}_r}]_{|_{{\chi_{{}_c} =0}}}=0\ ,
$$
which expresses the fact that, in order to reach equivalence with the original theory under the gauge fixing 
$q= c$, the EOM for the reduced theory, with Lagrangian $L_{{}_r}:= L_{|_{q= c}}$, must be supplemented with 
the imposition of the constraint $\chi_{{}_c} =0$. More general cases are discussed in the references cited 
above.
\section{Conclusions}
In this paper we have studied the reduction procedure which consists in the substitution of some fields by 
local functionals of the other fields. We give a formula, equation (\ref{theres}), which shows the origin of 
the possible mismatch between doing it at the level of the Lagrangian or at the level of the EOM. Along the 
way, we make contact with the method of Lagrange multipliers. We also discuss the conditions under which a 
symmetry of the original Lagrangian will yield a symmetry of the reduced Lagrangian. In the Noether case 
these conditions are essentially spelled out in equation (\ref{noet-cond}).

The particular case of auxilary variables is discussed in detail. For this kind of variables, the reduction 
is always consistent and the symmetries are always preserved. In particular, Noether symmetries are 
preserved as such. We show basically three uses for the auxiliary variables. First (section \ref{first}), 
they may bring an original non polynomial Lagrangian to a polynomial form; second (section \ref{second}), 
their may help to lower the order of the differential EOM; and third (section \ref{third}), they may be 
instrumental in closing off shell the algebra of the Noether symmetry generators.

Finally, we consider the case where the substitution of fields is made in the context of a gauge fixing 
procedure for a gauge theory. The essential lesson to be drawn is that the reduced theory may have lost some 
constraints that were present in the original theory. Once these constraints are reintroduced, 
consistency between the original and the reduced theory can be achieved.

\section*{Acknowledgements}
I thank Gary Gibbons and Joaqu\'{\i}m Gomis for very useful suggestions and for pointing out some
references. I also thank Juan Jos\'e L\'opez Villarejo for a useful comment. This work 
has been partially supported by MCYT FPA 2007-66665, CIRIT GC 2005SGR-00564, Spanish Consolider-Ingenio 2010 
Programme CPAN (CSD2007-00042).

\end{document}